\documentclass[aps,prb,twocolumn,showpacs,superscriptaddress,groupedaddress]{revtex4}
\usepackage{graphicx}
\usepackage{amssymb}
\usepackage{amsmath}
\usepackage{lmodern}
\usepackage{color}
\usepackage{hyperref}
\usepackage{epstopdf}
\usepackage{natbib}
\begin{document}

\title{Green's functions and DOS for some 2D lattices}

\author{Eugene Kogan}
\email{Eugene.Kogan@biu.ac.il}

\affiliation{Department of Physics, Bar-Ilan University, Ramat-Gan 52900, Israel}
\affiliation{Max-Planck-Institut f\"{u}r Physik komplexer Systeme,  Dresden 01187, Germany}

\author{Godfrey Gumbs}
\email{ggumbs@hunter.cuny.edu}

\affiliation{Department of Physics and Astronomy, Hunter College of the City
University of New York, 695 Park Avenue, New York, NY 10065, USA}

\date{\today}

\begin{abstract}
In this note we   present  the Green's functions and density of states
for the most frequently encountered 2D lattices: square, triangular, honeycomb, kagome, and Lieb lattice. Though the results are well known, we hope that their derivation  performed in a uniform way is of some pedagogical value.
\end{abstract}


\maketitle

\section{Introduction}

Fermionic lattice models are widely used not only as a
purely theoretical tool but also as a basis for investigation
and modelling of physical properties of real materials [\onlinecite{komnik}].
Despite their relative formal simplicity - the Hamiltonians of many of them can be written down as bilinears
of fermionic operators - analytical calculation of the lattice Green's functions can present substantial difficulties.

In general, the lattice Green's functions  of systems are
ubiquitous [\onlinecite{katsura,economou,varma}] in solid state physics, appearing in problems
of lattice vibrations, spin wave theory of magnetic systems,
localized oscillation modes at lattice defects, combinatorial
problems in lattices [\onlinecite{guttmann}], and flux calculations in lattice
percolation [\onlinecite{ziff}].
Lattice Green's functions  are also central to the theory of random walks on a lattice [\onlinecite{barber,hughes}], and to the
calculation of the effective resistance of resistor networks [\onlinecite{cserti}].

The lattice Green's functions  are of central importance for understanding the electronic behavior
of perfect crystalline solids. They also provide the basis for understanding the electronic properties of real,
imperfect crystalline solids, since the imperfections can be treated as a perturbation.
Lattice Green's functiona are also important for calculating RKKY interaction [\onlinecite{sherafati,satpathy,zare, gumbs}].

We will consider below several popular 2D lattices and calculate Green's functions and density of states (DOS).
The results are well known, still we think that derivation of these results performed in a uniform way and being presented in one place is of some pedagogical value.

In all cases we'll consider the models with the nearest neighbour hopping only,
the amplitude of the  hopping we'll take to be 1, so the Hamiltonian will be
\begin{eqnarray}
\label{hi}
H=\sum_{<ij>}c_i^{\dagger}c_j
\end{eqnarray}
where $c^{\dagger}$ and $c$ are electron creation and annihilation operators, and the summation in Eq. (\ref{hi}) is with respect to nearest neighbor pairs.
The period of the lattice we'll also take to be equal to 1.

\section{Square lattice}

In the wavevector representation the Hamiltonian for the square lattice is
\begin{eqnarray}
\label{hamiltonians}
H_{\Box}({\bf k})=\epsilon_{\Box}({\bf k})= 2\cos k_x+2\cos k_y.
\end{eqnarray}
The site diagonal matrix element of the Green's function is
\begin{eqnarray}
\label{gsq}
&&g_{\Box}(E)=\frac{1}{(2\pi)^2}\int_{-\pi}^{\pi}\int_{-\pi}^{\pi}\frac{dk_xdk_y}{E-\epsilon_{\Box}({\bf k})}\nonumber\\
&&=\frac{1}{(2\pi)^2}\int_{-\pi}^{\pi}\int_{-\pi}^{\pi}\frac{dk_xdk_y}{E-2\cos k_x-2\cos k_y}.
\end{eqnarray}
Employing the identity
\begin{eqnarray}
\label{identity}
\frac{1}{2\pi}\int_{-\pi}^{\pi}\frac{d\theta}{a+b\cos\theta}=\frac{\text{sign}(a)}{\sqrt{a^2-b^2}}\;\;\;\;\; (|a|>|b|)
\end{eqnarray}
to perform integration with respect to $k_y$, we obtain for $E<-4$ or $E>4$
\begin{eqnarray}
&&g_{\Box}(E)=\frac{1}{2\pi}\int_{-\pi}^{\pi}\frac{dk_x}{\sqrt{(E-2\cos k_x)^2-4}}\nonumber\\
&&=\frac{\text{sign}(E)}{2\pi}\int_{-1}^{1}\frac{dx}{\sqrt{(1-x^2)(x-c)(x-d)}},
\end{eqnarray}
where
$c=1+\frac{E}{2}$, $d=-1+\frac{E}{2}$.
For $E<-4$ we may  use the identity [\onlinecite{prudnikov}]
\begin{eqnarray}
\int_{b}^{a}\frac{dx}{\sqrt{(a-x)(x-b)(x-c)(x-d)}} \nonumber\\
=\frac{2}{\sqrt{(a-c)(b-d)}}K(k),
\end{eqnarray}
where $a>b>c>d$, $K$ is the complete elliptic integral of the first kind, and
\begin{eqnarray}
k=\sqrt{\frac{(a-b)(c-d)}{(a-c)(b-d)}}.
\end{eqnarray}
In our case $a=1$, $b=-1$; hence we obtain
\begin{eqnarray}
(a-b)(c-d)&=&4\nonumber\\
(a-c)(b-d)&=&\frac{E^2}{4}.
\end{eqnarray}
Thus we get for $E>4$
\begin{eqnarray}
g_{\Box}(E)=\frac{2}{\pi E}K\left(\frac{4}{E}\right).
\end{eqnarray}

\begin{figure}[h]
\includegraphics[width= .8\columnwidth]{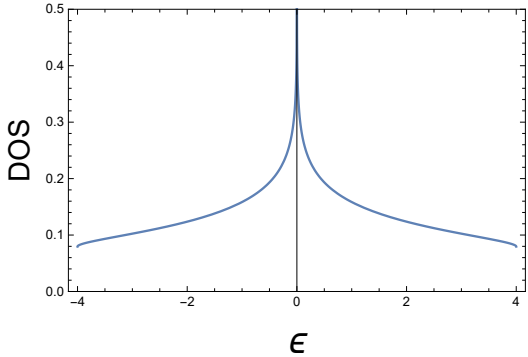}
\caption{DOS for the square lattice.}
 \label{square}
\end{figure}

We can analytically continue the Green function from the part of real axis
$E<-4$. For $-4<E<4$ we have $k>1$. Hence we should put $E=\epsilon+i0^+$, that is
$k$ acquires infinitesimal imaginary part $i0^+$, and we may use the identities [\onlinecite{economou}]
\begin{eqnarray}
\label{iden}
\text{Im}\left[K(k+i0^+)\right]&=&-\frac{1}{k} K\left(\sqrt{1-\frac{1}{k^2}}\right)\\
\label{iden2}
\text{Re}\left[K(k+i0^+)\right]&=&\frac{1}{k} K\left(\frac{1}{k}\right)
\end{eqnarray}
to  get [\onlinecite{economou}]
\begin{eqnarray}
\rho_{\Box}(\epsilon)=-\frac{1}{\pi}\text{Im}\left[g\left(\epsilon+i0^+\right)\right]
=\frac{1}{2\pi^2}K\left(\sqrt{1-\frac{\epsilon^2}{16}}\right).
\nonumber\\
\end{eqnarray}
The DOS is presented on Fig. \ref{square}

\section{Triangular lattice}
\label{tria}

Let ${\bf a}_1$, ${\bf a}_2$ and ${\bf a}_3$ are the three periods of the triangular lattice adding up to zero.
The Hamiltonian in the wave vector representation is
\begin{eqnarray}
\label{hamiltonian3}
&&H_{\triangle}({\bf k})=\epsilon_{\triangle}({\bf k})=2\cos k_1+2\cos k_2
+2\cos k_3,\nonumber\\
\end{eqnarray}
where $k_i\equiv {\bf a}_i\cdot {\bf k}$.
The site diagonal matrix element of the Green's function is ($k_1$ and $k_2$ are the components of ${\bf k}$ in the oblique coordinate system)
\begin{eqnarray}
\label{ggg}
&&g_{\triangle}(E)=\frac{1}{(2\pi)^2}\int_{-\pi}^{\pi}\int_{-\pi}^{\pi}\frac{dk_1dk_2}{E-\epsilon_{\triangle}({\bf k})}\\
&&=\frac{1}{(2\pi)^2}\int_{-\pi}^{\pi}\int_{-\pi}^{\pi}\frac{dk_1dk_2}{E-2\cos k_1-4\cos\left(k_2+\frac{k_1}{2}\right)\cos\left(\frac{k_1}{2}\right)}.\nonumber
\end{eqnarray}
(we used the transformation suggested in Ref. \onlinecite{hughes}).
Employing the identity (\ref{identity})
to perform integration with respect to $k_2$ we obtain for $E>6$ or $E<-2$
\begin{eqnarray}
&&g_{\triangle}(E)=\frac{\text{sign}(E)}{2\pi}\int_{-\pi}^{\pi}\frac{dk_1}{\sqrt{(E-2\cos k_1)^2-8-8\cos k_1}}\nonumber\\
&&=\frac{\text{sign}(E)}{8\pi}\int_{-\pi}^{\pi}\frac{dk_1}{\sqrt{(a-\cos k_1)(b-\cos k_1)}}\nonumber\\
&&=\frac{\text{sign}(E)}{4\pi}\int_{-1}^{1}\frac{dx}{\sqrt{(1-x^2)(a-x)(b-x)}},
\end{eqnarray}
where [\onlinecite{henyey}]
\begin{eqnarray}
\label{ab}
a&=&1+\frac{E}{2}+\sqrt{3+E}\nonumber\\
b&=&1+\frac{E}{2}-\sqrt{3+E}.
\end{eqnarray}
For $E>6$ we may  use the identity [\onlinecite{prudnikov}]
\begin{eqnarray}
\int_{d}^{c}\frac{dx}{\sqrt{(a-x)(b-x)(c-x)(x-d)}}  \nonumber\\
=\frac{2}{\sqrt{(a-c)(b-d)}}K(k),
\end{eqnarray}
where $a>b>c>d$,  and
\begin{eqnarray}
k=\sqrt{\frac{(a-b)(c-d)}{(a-c)(b-d)}}.
\end{eqnarray}
From Eq. (\ref{ab}) follows
\begin{eqnarray}
(a-b)(c-d)&=&4r\nonumber\\
(a-c)(b-d)&=&\frac{1}{4}(r-1)^3(r+3),
\end{eqnarray}
where
\begin{eqnarray}
r=\sqrt{3+E}.
\end{eqnarray}
Thus we get for $E>6$ [\onlinecite{muller,moritz}]
\begin{eqnarray}
\label{bt}
g_{\triangle}(E)=\frac{1}{\pi(r-1)^{3/2}(r+3)^{1/2}}
K\left(\frac{4r^{1/2}}{(r-1)^{3/2}(r+3)^{1/2}}\right). \nonumber\\
\end{eqnarray}

\begin{figure}[h]
\includegraphics[width= .8\columnwidth]{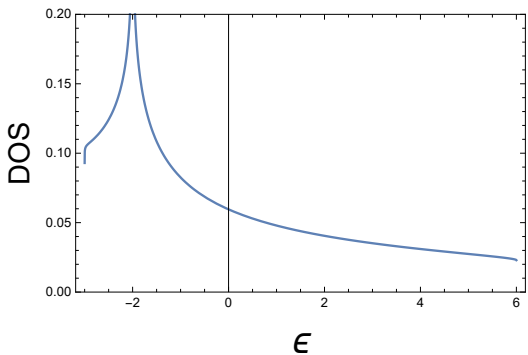}
\caption{DOS for the triangular lattice.}
 \label{triang}
\end{figure}

We can analytically continue the Green function from the part of real axis
$E>6$. For $-2<E<6$ we have $k>1$. Hence we should put $E=\epsilon+i0^+$, that is
$k$ acquires infinitesimal imaginary part $i0^+$, we may use the identities (\ref{iden}),
(\ref{iden2})
and take into account that
\begin{eqnarray}
1-\frac{1}{k^2}=\frac{(3-r)(r+1)^3}{16r}.
\end{eqnarray}
For $-3<E<2$ the value of $k$ become imaginary. We may use the identity [\onlinecite{erdelyi}]
\begin{eqnarray}
K(ik)=\kappa'K(\kappa),
\end{eqnarray}
where
\begin{eqnarray}
\kappa=\frac{k}{\sqrt{k^2+1}},\;\;\;
\kappa'=
\frac{1}{\sqrt{k^2+1}}.
\end{eqnarray}
In our case
\begin{eqnarray}
\kappa=\sqrt{\frac{4r}{(3-r)(r+1)^{3/4}}},
\end{eqnarray}
and we reproduce DOS for the triangular lattice [\onlinecite{muller}]
\begin{eqnarray}
\rho_{\triangle}(\epsilon)=-\frac{1}{\pi}\text{Im}\left[g\left(\epsilon+i0^+\right)\right]
=\frac{1}{\pi^2\sqrt{z_0}}K\left(\sqrt{\frac{z_1}{z_0}}\right),\nonumber\\
\end{eqnarray}
where
\begin{eqnarray}
z_0=\left\{\begin{array}{lll} \frac{(3-r)(r+1)^3}{4} & \text{for} & -3\leq\epsilon\leq -2\\
                             4r & \text{for} & -2\leq\epsilon\leq 6
            \end{array}\right.,
\end{eqnarray}
\begin{eqnarray}
z_1=\left\{\begin{array}{lll} 4r & \text{for} & -3\leq\epsilon\leq -2\\
                              \frac{(3-r)(r+1)^3}{4} & \text{for} & -2\leq\epsilon\leq 6
            \end{array}\right.
\end{eqnarray}
The DOS is presented on Fig. \ref{triang}.

\section{Honeycomb lattice}

The honeycomb  lattice can be considered as a triangular lattice
with a basis of two lattice points.The tight binding Hamiltonian for the electrons  is
\begin{eqnarray}
\label{ham4}
\hat{H}=\sum_{\bf n}a_{\bf n}^{\dagger}(b_{{\bf n}+\boldsymbol{\delta}_1}+b_{{\bf n}+\boldsymbol{\delta}_2}+b_{{\bf n}+\boldsymbol{\delta}_3})+\text{H.c},
\end{eqnarray}
where
$\boldsymbol{\delta}_1,
\boldsymbol{\delta}_2,\boldsymbol{\delta}_3$ are the vectors connecting an atom with its nearest neighbors.

Going to wave vector representation we obtain
\begin{eqnarray}
\label{hamiltonian}
\hat{H}=\sum_{\bf k}\Psi^{\dagger}_{\bf k}\hat{H}({\bf k})\Psi_{\bf k},
\end{eqnarray}
where
\begin{eqnarray}
\label{hamiltonian2a}
\hat{H}({\bf k})=\left(\begin{array}{cc} 0  & \sum_ie^{-i{\bf k}\cdot\boldsymbol{\delta}_i}  \\
     \sum_ie^{i{\bf k}\cdot\boldsymbol{\delta}_i}  &  0  \end{array}\right).
\end{eqnarray}
Taking into account that
$\boldsymbol{\delta}_1-\boldsymbol{\delta}_2={\bf a}_3$,
$\boldsymbol{\delta}_2-\boldsymbol{\delta}_3={\bf a}_1$,
$\boldsymbol{\delta}_3-\boldsymbol{\delta}_1={\bf a}_2$,
we obtain the spectrum as
\begin{eqnarray}
\epsilon_{\nu}({\bf k})=\nu\sqrt{X_{\bf k}},
\end{eqnarray}
where $\nu=\pm 1$, and
\begin{eqnarray}
\label{x0}
X_{\bf k}\equiv 2(\cos k_1+\cos k_2+\cos k_3)+3.
\end{eqnarray}

\begin{figure}[h]
\includegraphics[width= .8\columnwidth]{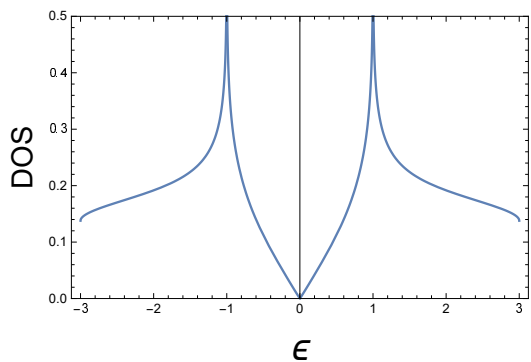}
\caption{DOS for the honeycomb lattice.}
 \label{honey}
\end{figure}

The dispersion law can be written in the form
\begin{eqnarray}
\epsilon_{\nu}({\bf k})=\nu\sqrt{3+\epsilon_{\triangle}({\bf k})}.
\end{eqnarray}
The Green's function is
\begin{eqnarray}
g_H(E)=\frac{1}{2\pi^2}\int_{-\pi}^{\pi}dk_1\int_{-\pi}^{\pi}dk_2
\frac{E}{E^2-3-\epsilon_{\triangle}(k)}.
\end{eqnarray}
Thus we obtain
\begin{eqnarray}
\label{bth}
g_{H}(E)=2Eg_{\triangle}(E^2-3).
\end{eqnarray}
For the density of states we obtain
\begin{eqnarray}
\label{hc}
\rho_H=2|\epsilon|\rho_{\triangle}(\epsilon^2-3).
\end{eqnarray}
Actually, Eqs. (\ref{bth}) and (\ref{hc}) become completely obvious after we square the Hamiltonian (\ref{hamiltonian2a}) to get
\begin{eqnarray}
\label{milton}
\hat{H}^2({\bf k})=\left(\begin{array}{cc} \epsilon_{\triangle}({\bf k})+3  & 0  \\
0  & \epsilon_{\triangle}({\bf k})+3  \end{array}\right).
\end{eqnarray}
The DOS is presented on Fig. \ref{honey}

\section{Kagome lattice}

The kagome lattice can be considered as a triangular lattice
(which we consider to be identical to the lattice introduced above for the honeycomb lattice)
with a basis of three lattice points.
The Hamiltonian is
\begin{eqnarray}
\label{ham0}
&&\hat{H}=\sum_{\bf n}\left[a_{\bf n}^{\dagger}(b_{{\bf n}+\boldsymbol{\delta}_3}+b_{{\bf n}-\boldsymbol{\delta}_3})\right.\\
&&+\left.a_{\bf n}^{\dagger}(c_{{\bf n}+\boldsymbol{\delta}_2}+c_{{\bf n}-\boldsymbol{\delta}_2})
+b_{{\bf n}+\boldsymbol{\delta}_3}^{\dagger}(c_{{\bf n}+\boldsymbol{\delta}_1}+c_{{\bf n}-\boldsymbol{\delta}_1})+\text{H.c}\right],\nonumber
\end{eqnarray}
where
$\boldsymbol{\delta}_i={\bf a}_i/2$ are the vectors connecting the nearest neighbors.

Going to wave vector representation we obtain
\begin{eqnarray}
\label{hamiltonianb}
\hat{H}=\sum_{\bf k}\Psi^{\dagger}_{\bf k}\hat{H}({\bf k})\Psi_{\bf k},
\end{eqnarray}
where
\begin{eqnarray}
\label{hamiltonian2}
\hat{H}({\bf k})=2\left(\begin{array}{ccc} 0  & \cos \left(\frac{k_3}{2}\right) &  \cos \left(\frac{k_2}{2}\right)  \\
     \cos \left(\frac{k_3}{2}\right)  &  0 &  \cos \left(\frac{k_1}{2}\right)  \\
       \cos \left(\frac{k_2}{2}\right)  &  \cos \left(\frac{k_1}{2}\right)  & 0
       \end{array}\right).
\end{eqnarray}
The spectrum is found from the equation
\begin{eqnarray}
\text{det}\left(\epsilon\hat{I}-\hat{H}\right)=0.
\end{eqnarray}
Calculating the determinant and taking into account that $k_1+k_2+k_3=0$ we may   present the dispersion equation as
\begin{eqnarray}
[\epsilon({\bf k})+2][\epsilon^2({\bf k}) -2\epsilon({\bf k})-X_{\bf k}+1] =0.
\end{eqnarray}
Thus we have a flat band $\epsilon_F({\bf k})=-2$, and two dispersive bands
\begin{eqnarray}
\epsilon_{\nu}({\bf k})=1+\nu\sqrt{X_{\bf k}}.
\end{eqnarray}

The Green's function is
\begin{eqnarray}
\label{ggk}
&&g_K(E)=\frac{1}{(2\pi)^2}\int_{-\pi}^{\pi}\int_{-\pi}^{\pi}\frac{dk_1dk_2}{E+2}\\
&&+\frac{1}{2\pi^2}\int_{-\pi}^{\pi}dk_1\int_{-\pi}^{\pi}dk_2
\frac{E-1}{E^2-2E-2-\epsilon_{\triangle}(k)}.\nonumber
\end{eqnarray}
Comparing Eqs. (\ref{ggk}) and (\ref{ggg}) we obtain
\begin{eqnarray}
\label{bth0}
g_K(E)=\frac{1}{E+2}+2(E-1)g_{\triangle}[(E-1)^2-3]
\end{eqnarray}
Hence
\begin{eqnarray}
\label{hc0}
\rho_K=\delta(\epsilon+2)+2|\epsilon-1|\rho_{\triangle}[(\epsilon-1)^2-3].
\end{eqnarray}

\begin{figure}[h]
\includegraphics[width= .8\columnwidth]{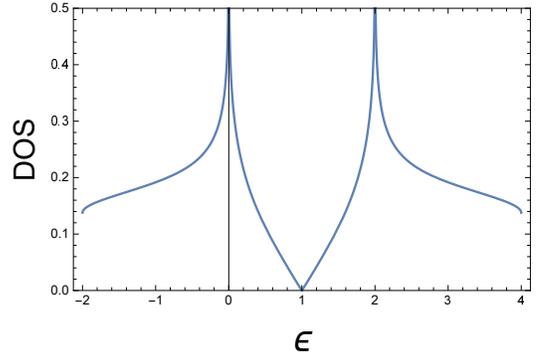}
\caption{DOS for the kagome lattice. (The $\delta$-function peak is omitted.)}
 \label{kag}
\end{figure}

We see, that apart from the $\delta$-function peak, the DOS for the Kagome lattice is the DOS for the honeycomb lattice shifted by 1.
The DOS is presented on Fig. \ref{kag}

\section{Lieb lattice}
\label{lieb0}

\begin{figure}[h]
\includegraphics[width= .8\columnwidth]{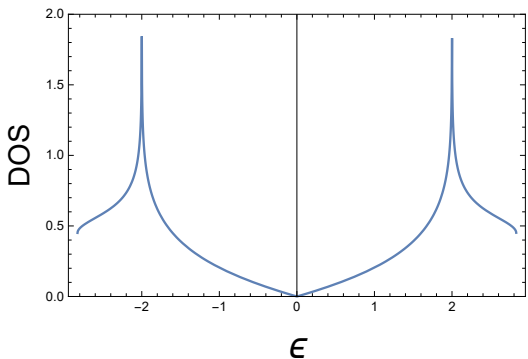}
\caption{DOS for the Lieb lattice. (The $\delta$-function peak is omitted.)}
 \label{liib}
\end{figure}

Consider now the  Lieb lattice with the Hamiltonian
\begin{eqnarray}
\label{ham16}
\hat{H}({\bf k}) =2\left(\begin{array}{ccc}
0 & \cos\left(\frac{k_x}{2}\right) & 0 \\
\cos\left(\frac{k_x}{2}\right) & 0 &
\cos\left(\frac{k_y}{2}\right)\\
0 &  \cos\left(\frac{k_y}{2}\right) & 0 \end{array}\right).
\end{eqnarray}
The spectrum of the Hamiltonian (\ref{ham16}) is
\begin{eqnarray}
\epsilon_F({\bf k})=0,\quad
\epsilon_{\nu}({\bf k})=2\nu\sqrt{\cos^2\left(\frac{k_x}{2}\right) +
\cos^2\left(\frac{k_y}{2}\right)}.\nonumber\\
\end{eqnarray}
The Green's function is
\begin{eqnarray}
\label{ggk2}
&&g_L(E)=\frac{1}{(2\pi)^2}\int_{-\pi}^{\pi}\int_{-\pi}^{\pi}\frac{dk_1dk_2}{E}\\
&&+\frac{1}{2\pi^2}\int_{-\pi}^{\pi}dk_1\int_{-\pi}^{\pi}dk_2
\frac{2E}{E^2-4-\epsilon_{\Box}(k)}.\nonumber
\end{eqnarray}
Comparing Eqs. (\ref{ggk2}) and (\ref{gsq}) we obtain
\begin{eqnarray}
\label{bth2}
g_L(E)=\frac{1}{E}+2Eg_{\Box}\left(E^2-4\right).
\end{eqnarray}
Hence
\begin{eqnarray}
\label{hc2}
\rho_K&=&\delta(\epsilon)+2|\epsilon|\rho_{\Box}(\epsilon^2-4)\nonumber\\
&=&\delta(\epsilon)+\frac{|\epsilon|}{\pi^2}K\left(\sqrt{1-\frac{(\epsilon^2-4)^2}{16}}\right).
\end{eqnarray}
The DOS is presented on Fig. \ref{liib}

\section{Conclusions}

To summarize, we have expressed lattice Green's functions
and density of states for triangular and square lattice through the complete elliptic integral of the first kind in a uniform way. The Green's functions were calculated analytically as functions of energy on the part of real axis, and then continued analytically on the whole complex plane.

We connected lattice Green's functions
and density of states for honeycomb and kagome lattices with those of the triangular lattice, and Green's functions
and density of states for Lieb lattice with those of square lattice.
Exact
expressions for lattice density of states ought to be useful in
dynamical mean field theory calculations [\onlinecite{13}].

We have also shown that the well known results for the wave functions of electrons in the vicinity of the Dirac points in the honeycomb lattice are not connected with the nearest-neighbor hopping approximation, but follow from the symmetry of the model.

\begin{acknowledgments}

The work on this paper started  during E.K. visit to
Max-Planck-Institut f\"{u}r Physik komplexer Systeme in December of 2019 and January of 2020.
E.K. cordially thanks the Institute for the hospitality extended to him during
that and all his previous visits.

\end{acknowledgments}

\begin{appendix}

\section{Why the wave functions of electrons in honeycomb lattice are what they are?}
\label{hon}

The results for the wave functions of electrons in the vicinity of the Dirac points
in the honeycomb lattice (Eq. (\ref{hc5b}) and
(\ref{hc6}) below) are   well known [\onlinecite{castro}].
In this appendix we want to emphasise the fact that these results are not connected, as it is sometimes erroneously stated, with the nearest-neighbor hopping approximation, but follow from the symmetry of the model.

Looking at Eq. (\ref{x0}) we understand that to find minimal value of $X_{\bf k}$ we should find minimal value of the sum of cosines of three numbers
which add up to zero.
 This value  is equal to $-3/2$ and is achieved when two of the numbers are $2\pi/3$ and the third number is $-4\pi/3$ (the point ${\bf K}$), or
when two of the numbers are $-2\pi/3$ and the third number is $4\pi/3$ (the point ${\bf K}'$). The bands touch each other at these points.

The wave functions can be presented as
\begin{eqnarray}
\label{hc5}
\Psi_{\nu}({\bf k})=\frac{1}{\sqrt{2}}\left(\begin{array}{l}e^{-i\widetilde{\theta}_{\bf k}/2} \\
\nu e^{i\widetilde{\theta}_{\bf k}/2}\end{array}\right),
\end{eqnarray}
where
\begin{eqnarray}
\widetilde{\theta}_{\bf k}
=\text{Arg}\left(\sum_ie^{i\boldsymbol{\delta}_i\cdot{\bf k}}\right).
\end{eqnarray}

Making substitution
\begin{eqnarray}
{\bf k}={\bf K}+{\bf q},
\end{eqnarray}
expanding with respect to ${\bf q}$ and keeping only the linear terms we obtain
\begin{eqnarray}
\label{x3}
X_{\bf q}=\frac{1}{2}\sum_i({\bf a}_i\cdot{\bf q})^2=\frac{3}{4}q^2.
\end{eqnarray}
Thus we have conic point in the spectrum.

In the same approximation
\begin{eqnarray}
\label{hc10}
\sum_ie^{i\boldsymbol{\delta}_i\cdot{\bf k}}=\sum_ie^{i\boldsymbol{\delta}_i\cdot{\bf K}}
\left(\boldsymbol{\delta}_i\cdot{\bf q}\right)=({\bf n}+i{\bf m})\cdot{\bf q},
\end{eqnarray}
where
\begin{eqnarray}
{\bf n}=\sum_i\boldsymbol{\delta}_i\cos(\boldsymbol{\delta}_i\cdot{\bf K})
,\;\;{\bf m}=\sum_i\boldsymbol{\delta}_i\sin({\bf K}\boldsymbol{\delta}_i).
\end{eqnarray}
One can easily check up that
\begin{eqnarray}
\label{perp}
{\bf n}\perp{\bf m},\;\; |{\bf n}|=|{\bf m}|.
\end{eqnarray}
 Hence $\widetilde{\theta}_{\bf q}=\theta_{\bf q}$,
where $\theta_{\bf q}$
is  the polar angle of ${\bf q}$, the $X$ axis being chosen in the direction of ${\bf n}$.
If we chose
\begin{eqnarray}
\boldsymbol{\delta}_1=\frac{1}{2}\left(\frac{1}{\sqrt{3}},1\right), \;\;\;
\boldsymbol{\delta}_2=\frac{1}{2}\left(\frac{1}{\sqrt{3}},-1\right), \nonumber\\
\boldsymbol{\delta}_3=-\left(\frac{1}{\sqrt{3}},0\right),\;\;\;
{\bf K}=\left(\frac{2\pi}{\sqrt{3}},\frac{2\pi}{3}\right),
\end{eqnarray}
then $X$ axis is in the direction of $\boldsymbol{\delta}_2$.
Hence Eq. (\ref{hc5}) can be written as
\begin{eqnarray}
\label{hc5b}
\Psi_{\nu,{\bf K}}({\bf q})=\frac{1}{\sqrt{2}}\left(\begin{array}{l}e^{-i\theta_{\bf q}/2} \\
\nu e^{i\theta_{\bf q}/2}\end{array}\right).
\end{eqnarray}
For the ${\bf K}'$ point, the wavefunctions are obtained from those in Eq. (\ref{hc5b}) by permutation of the sublattices
\begin{eqnarray}
\label{hc6}
\Psi_{\nu,{\bf K}'}({\bf q})=\Psi_{\nu,{\bf K}}^*({\bf q}).
\end{eqnarray}

The natural question appears:  are Eqs. (\ref{hc5b}) and (\ref{hc6})
connected to the simplest possible tight-binding model we have used?
The answer is: No, these equations are general, and follow from the symmetry of the problem.
In fact, the general tight-binding Hamiltonian  for the honeycomb lattice is
\begin{eqnarray}
\label{ham}
\hat{H} =
\left(\begin{array}{cc}
\sum_{\bf a} t'({\bf a})e^{i{\bf k\cdot a}} & \sum_{\bf a}t({\bf a}+\boldsymbol{\delta})e^{i{\bf k\cdot}({\bf a}+\boldsymbol{\delta})}\\
\sum_{\bf a}t^*({\bf a}+\boldsymbol{\delta})e^{-i{\bf k\cdot}({\bf a}+\boldsymbol{\delta})} &  \sum_{\bf a}t'({\bf a})e^{i{\bf k\cdot a}} \end{array}\right),\nonumber\\
\end{eqnarray}
where the summation is with respect to all  lattice vectors ${\bf a}$, and $\boldsymbol{\delta}$ is some arbitrary, but fixed $\boldsymbol{\delta}_i$.
The selection rule for matrix elements
[\onlinecite{landau}] gives
\begin{eqnarray}
\label{zero}
\sum_{\bf a}t({\bf a}+{\bf \delta})e^{i{\bf K\cdot}({\bf a}+\boldsymbol{\delta})}=0.
\end{eqnarray}
In fact, we are dealing with the product of two functions. The function $t({\bf a}+\boldsymbol{\delta})$ realizes the unit representation
of the point symmetry group $C_{3}$ (the full symmetry group of the inter--sublattice hopping is $C_{3v}$, but the restricted symmetry is enough to prove the cancelation).
As far as the function $e^{i{\bf K\cdot}({\bf a}+\boldsymbol{\delta})}$ is concerned, rotation of the lattice by the angle $2\pi/3$,
say anticlockwise,
 is equivalent to rotation of the vector ${\bf K}$ in the opposite direction,
 that is to substitution of the three equivalent corners of the Brilluoin zone. Thus  the exponent $e^{i{\bf K\cdot}({\bf a}+\boldsymbol{\delta})}$
realizes  $x-iy$ representation of the group $C_{3}$. Because each of multipliers in Eq. (\ref{zero}) realizes different irreducible representation of the symmetry group, the matrix element is equal to zero.
Simply speaking, at a point {\bf K} the sublattices become decoupled, and this explains the degeneracy of the electron states in this point (these points) or, in other words, merging of the two branches of the single Brilouin zone.

Hence,  for a general tight-binding model, the nondiagonal matrix element of the Hamiltonian  (\ref{ham})
is a linear function of ${\bf q}$:
$\left({\bf c}+i{\bf d}\right)\cdot{\bf q}$,
invariant with respect to rotation of ${\bf q}$ by $2\pi/3$ up to a ${\bf q}$ independent constant.
It immediately leads to the demands
$|{\bf c}|=|{\bf d}|$, and ${\bf c}\perp{\bf d}$ and we recover Eq. (\ref{hc5b}).
Equation (\ref{hc6}) follows from Eq. (\ref{hc5b}) because of invariance of the system with respect to mirror reflections.

\end{appendix}



\begin{thebibliography}{99}

\bibitem{komnik} A. Komnik and S. Heinze, Analytical results for the Green's functions of lattice fermions, Phys. Rev. B {\bf 96}, 155103-1-155103-13 (2017); https://doi.org/10.1103/PhysRevB.96.155103.

\bibitem{varma} V. K. Varma and H. Monien, Lattice Green’s functions for kagome, diced, and hyperkagome lattices, Phys. Rev. E {\bf 87}, 032109-1-032109-4 (2013); https://doi.org/10.1103/PhysRevE.87.032109.

\bibitem{katsura}  S. Katsura, T. Morita, S. Inawashiro, T.
Horiguchi, and Y. Abe, Lattice Green's Function. Introduction, J. Math. Phys. {\bf 12}, 892-895 (1971); https://doi.org/10.1063/1.1665662.

\bibitem{economou} E. N. Economou, {\it Green's Functions in Quantum Physics, 3 ed.}, (Springer-Verlag, Berlin-Heidelberg, 2006).

\bibitem{guttmann} A. J. Guttmann and T. Prellberg, Staircase polygons, elliptic integrals, Heun functions, and lattice Green functions, Phys. Rev. E {\bf 47}, R2233-R2236
(1993); https://doi.org/10.1103/PhysRevE.47.R2233.

\bibitem{ziff} R. M. Ziff, Flux to a trap, J. Stat. Phys. {\bf 65}, 1217-1233 (1991); https://doi.org/10.1007/BF01049608.

\bibitem{barber} M. N. Barber and B. W. Ninham,   {\it Random and Restricted Walks} (New York: Gordon and Breach, 1970).

\bibitem{hughes} B. D. Hughes, {\it Random Walks and Random Environments. Volume 1: Random Walks} (Oxford: Clarendon, 1995).

\bibitem{cserti} J. Cserti, Application of the lattice Green's function for calculating the resistance of an infinite network of resistors,  Am. J. Phys. {\bf 68} 896-906 (2000); https://doi.org/10.1119/1.1285881.

\bibitem{sherafati} M. Sherafati, S. Satpathy, RKKY Interaction in Graphene from Lattice Green's Function, Phys. Rev. B {\bf 83}, 165425-1-165425-8 (2011); https://doi.org/10.1103/PhysRevB.83.165425

\bibitem{satpathy} F. Parhizgar, M. Sherafati, Reza Asgari, S. Satpathy, Ruderman-Kittel-Kasuya-Yosida interaction in biased bilayer graphene, Phys. Rev. B {\bf 87}, 165429-1-165429-11 (2013); https://doi.org/10.1103/PhysRevB.87.165429.

\bibitem{zare} M. Zare, RKKY plateau in zero- and one-dimensional triangular Kagome lattice models, arXiv:1909.03406.

\bibitem{gumbs} O. Roslyak, G. Gumbs, A. Balassis, H. Elsayed, Effect of magnetic field and chemical potential on the RKKY interaction in the $\alpha-{\cal T}^3$
lattice, arXiv:2006.15447.


\bibitem{prudnikov} A. P. Prudnikov, Yu. A. Brychkov, and O. I. Marichev, {\it Integrals and
Series, Vol. 2},  (Gordon and Breach Science Publishers, Amsterdam,
1986).


\bibitem{henyey} F. S. Henyey and V. Seshadri, On the number of distinct sites visited in 2D lattices, J. Chem. Phys. {\bf 76}, 5530-5534 (1982); https://doi.org/10.1063/1.442908.

\bibitem{muller} Th. Hanisch, G. S. Uhrig, and E. Muller-Hartmann, Lattice dependence of saturated ferromagnetism in the Hubbard model, \prb {\bf 56}, 13960-13982 (1997); https://doi.org/10.1103/PhysRevB.56.13960.

\bibitem{moritz} B. Moritz and W. Schwalm, Triangle lattice Green functions for vector fields, J. Phys. A: Math. Gen. {\bf 34}, 589-602 (2001); https://doi.org/10.1088/0305-4470/34/3/317.

\bibitem{erdelyi} A. Erdelyi (ed.), {\it Higher Transcendental Functions, Vol. II}, (McGraw-Hill Book Company, Inc., 1985).




\bibitem{13} A. Georges, G. Kotliar, W. Krauth, M. Rozenberg, Dynamical mean-field theory of strongly correlated fermion systems and the limit of infinite dimensions, Rev. Mod. Phys. {\bf 68}(1), 13-125 (1996); doi:10.1103/RevModPhys.68.13.

\bibitem{castro}  F. Guinea, N. M. R. Peres, K. S. Novoselov, and A. K. Geim, The electronic properties of graphene
A. H. Castro Neto,
Rev. Mod. Phys. {\bf 81}, 109-162 (2009); https://doi.org/10.1103/RevModPhys.81.109.

\bibitem{landau} L. D. Landau and E. M. Lifshitz, {\it Quantum Mechanics}, (Pergamon Press, 1991).


\end{thebibliography}
\end{document}